\documentstyle[epsfig]{article}
\begin{document}
\newcommand{\beq}{\begin{equation}}
\newcommand{\eeq}{\end{equation}}
\newcommand{\bea}{\begin{eqnarray}}
\newcommand{\eea}{\end{eqnarray}}
\newcommand{\cl}[1]{\begin{center} {#1} \end{center}}
\newcommand{\dr}{d^3r}
\newcommand{\dxp}{d^3x^{'}}
\newcommand{\dxpp}{d^3x^{''}}
\newcommand{\ddp}{\frac{d^3p}{(2 \pi)^3}}
\newcommand{\dpj}{\frac{d^3p_1}{(2 \pi)^3}}
\newcommand{\dpp}{\frac{d^3p^{'}}{(2 \pi)^3}}
\newcommand{\dpjp}{\frac{d^{3}p_{1}^{'}}{(2 \pi)^3}}
\newcommand{\dl}{\frac{d^3l}{(2 \pi)^3}}
\newcommand{\p}{{\bf p}}
\newcommand{\pj}{{\bf p_{1}}}
\newcommand{\pp}{{\bf p^{'}}}
\newcommand{\pjp}{{\bf p_{1}^{'}}}
\newcommand{\k}{{\bf k}}
\cl{\Large{{\bf Nonlinearities of the Vlasov equation in the spinodal
region}}} 
\vskip 1truecm

\cl{     P. Bo\.{z}ek \footnote{e-mail~: bozek@quark.ifj.edu.pl} }
 
\vskip 1truecm
\cl{ Yukawa Institute for Theoretical Physics, Kyoto University,
Kyoto 606-01 , Japan}
\cl{ and}
\cl{ Institute of Nuclear Physics, 31-342 Krak\'{o}w, Poland}
     
%\date{}

\begin{abstract}
We estimate the importance of the nonlinear terms in the Vlasov
equation for the development of the unstable modes.
The results allow to identify the region of wavelength where the
linear evolution is justified.
\end{abstract}

\vspace{0.5cm}


\vspace{0.5cm}

\noindent 
Keywords~: Vlasov equation, spinodal decomposition, unstable modes,
mode-mode coupling

\vspace{3cm}

%%%%%\newpage

Recently  the Vlasov equation has been applied  for the description 
of the dynamics of the fragment formation in the nuclear matter
unstable against the spinodal decomposition \cite{c1,ch1,ch2}.
In the linear approximation a perturbation of the translationally
invariant distribution is unstable. Unstable modes will
develop exponentially in time leading to the  formation of density
inhomogeneities and fragmentation \cite{c1,he}. Since the growth rate
depends on  the 
wave-vector, the wave-vector with the largest growth rate
$\Gamma_{k}$ will dominate
at large times. One expects the formation of fragments of the size
corresponding to the most unstable modes $1/k$ \cite{he}.
If the linear regime can be trusted in the development of the
spinodal decomposition then the fragment size distribution should show 
an excess of fragments of intermediate sizes.
The fragment size distribution would be different if at the fragmentation
time
many different modes are excited due to the nonlinear evolution of the
Vlasov equation. Of course, in realistic simulations initial
fluctuations and a noise term must be taken into account. This
determines the strength by which different modes of the linearized
Vlasov equation would be excited.

The nonlinearities in the Vlasov equation are non-negligible
at large times as the deviation of the unstable modes from equilibrium 
becomes important. The existing estimates of this effect are based on
the numerical results for the 2-Dimensional Vlasov equation
\cite{ch1,ch2}. In this letter we shall estimate the importance of the 
nonlinear effects in the Vlasov equation analytically.
We shall study the case where at initial time $t=0$  
one or two unstable modes are perturbed. One unstable mode engenders
nonlinear effects at double wave-vector. Two unstable modes can lead
to nonlinearities also for small wave-vector. 
The estimate will allow to define the regime of wave-vectors where the
linear approximation can be valid.

We start with the usual Vlasov equation~:
\beq
\frac{\partial f(t,x,p)}{\partial t} + {\bf v}{\bf \nabla_x}
f(t,x,p) -
{\bf \nabla_x} U(x) {\bf \nabla_p} f(t,x,p) = 0 \ ,
\eeq
where $U(x)$ is the density dependent mean field. In the following we
take a Skyrme type mean field potential folded with a Gaussian~:
\beq
U(x)=\int d^3 y \ g(x-y) \  \Big(A
\frac{\rho(y)}{\rho_0}+B\Big(\frac{\rho(y)}{\rho_0}
\Big)^\sigma \Big) \ .
\eeq
Expanding to second order in $\delta f(t,x,p)$,
 the deviation
 from the homogeneous solution, and Fourier transforming in x,
$\delta f(t,k,p)=\int d^3 x \ \delta f(t,x,p) e^{-i{\bf x k}}$, 
we obtain~:
\bea
\label{nn}
\frac{\partial f(t,k,p)}{\partial t} & + & i {\bf k v} \delta f(t,k,p)
-i \frac{\delta U_k}{\delta \rho} {\bf k v} \frac{\partial
n_0}{\partial \epsilon} \delta \rho(t,k) \nonumber \\
& - & \frac{i}{2}\frac{\delta^2 U_k}{\delta \rho^2} {\bf k v}
\frac{\partial n_0}{\partial \epsilon} \int
\frac{d^3 l}{(2 \pi)^3} \delta \rho(t,k-l)  \ \delta \rho(t,l) 
\nonumber \\
 & - & \frac{i}{2} \int
\frac{d^3 l}{(2 \pi)^3} \frac{\delta U_{k-l}}{\delta \rho} \delta 
\rho(t,k-l)  \ ({\bf k - l}) {\bf \nabla}_p \delta f(t,k,p) =0 \ ,
\eea
where $n_0$ is the equilibrium momentum distribution.
The first three terms of the above equation are the linear
approximation to the Vlasov equation. The remaining two terms are the
second order expansion in $\delta f$ and represent a mode-mode coupling 
for the modes of the linear equation. The actual solution of the
second order
equation is as difficult as for the full Vlasov equation. However, to
make an estimate of the nonlinear effects it is enough to take for 
$\delta f(t,k,p)$ and $\delta \rho(t,k)$ in the nonlinear terms only
the contribution from the unstable modes to the evolution of an initial phase
space density perturbation at $t=0$~:
\beq
\delta f^{+}(t,k,p)= \frac{\delta U_k}{\delta \rho} \frac{\partial
n_0} {\partial \epsilon} \frac{{\bf kv}}{{\bf kv}-i \Gamma_l} \delta 
\rho^{+}(k) e^{\Gamma_k t} ,
\eeq
and 
\beq
\label{f1}
\delta \rho^{+}(t,k)=\delta \rho^{+}(k) \Theta(k_{max} -|k|) e^{\Gamma_k t} \ ,
\eeq
with $k_{max}$ being the maximal wave-vector with imaginary frequency \cite{c1}
and 
\beq
\label{f2}
\delta \rho^{+}(k)=\int \frac{d^3 p}{(2 \pi)^3} \frac{\delta f(0,k,p)}{{\bf
kv} - i \Gamma_k} \ .
\eeq

Inserting the  forms (\ref{f1}) and (\ref{f2}) in the mode-mode
coupling term in eq. (\ref{nn}) and taking the 
 one-sided Fourier transform ~:
\beq
\delta f(\omega,k,p) = \int_0^{\infty} dt \ \delta  f(t,k,p) e^{i \omega t}\ ,
\eeq
with $Im \ \omega > 2 \Gamma_{max}$,
one obtains~:
\bea
\label{p2}
i({\bf kv} - \omega) \delta f(\omega, k, p) & - &i \frac{\delta
U_k}{\delta \rho} {\bf k v} \frac{\partial n_0}{\partial \epsilon}
\delta \rho(\omega,k) \nonumber \\  & =  & \delta f(0,k,p) - \frac{i}{2}
\frac{\delta^2 U_k}{\delta \rho^2} {\bf kv } \frac{\partial
n_0}{\partial \epsilon} \int \frac{d^3 l}{(2 \pi)^3} \delta
\rho^{+}(k-l) \delta \rho^{+}(l) \nonumber \\
& & \frac{1}{i \omega
+\Gamma_{k-l}+\Gamma_l}
\nonumber \\
& - & \frac{i}{2} \int \frac{d^3 l}{(2 \pi)^3} \frac{\delta
U_{k-l}}{\delta \rho} \frac{\delta
U_{l}}{\delta \rho} \delta
\rho^{+}(k-l) \delta \rho^{+}(l) \frac{1}{i \omega
+\Gamma_{k-l}+\Gamma_l} 
\nonumber \\ & &  ({\bf k - l}) {\bf \nabla_p}
\Big( \frac{{\bf l v}}{{\bf l v }- i \Gamma_l}\frac{\partial
n_0}{\partial \epsilon} \Big) \ .
\eea
Dividing by ${\bf k}{\bf v} - \omega$ and integrating over $p$ an
equation for the density perturbation is obtained~:
\bea
\label{lll}
\delta \rho(\omega, k)  & = & \frac{ -i G(\omega,k)}{\epsilon(\omega,k)}
\nonumber \\
 & - & \frac{1}{2 \epsilon(\omega,k)} 
\frac{\delta^2 U_k}{\delta \rho^2} \int \frac{d^3 p}{(2
\pi)^3}\frac{\bf kv }{{\bf kv} - \omega}
 \frac{\partial
n_0}{\partial \epsilon} \int \frac{d^3 l}{(2 \pi)^3} \delta
\rho^{+}(k-l) \delta \rho^{+}(l) \frac{1}{i \omega
+\Gamma_{k-l}+\Gamma_l}
\nonumber \\
& - & \frac{1}{2 \epsilon(\omega,k)} \int \frac{d^3 l}{(2 \pi)^3} \frac{\delta
U_{k-l}}{\delta \rho} \frac{\delta
U_{l}}{\delta \rho} \delta
\rho^{+}(k-l) \delta \rho^{+}(l)
 \frac{1}{i \omega
+\Gamma_{k-l}+\Gamma_l} \nonumber \\
& &  \int \frac{d^3 p}{(2 \pi)^3}
\frac{1}
{{\bf kv} -
\omega}({\bf k - l}) {\bf \nabla_p}
\Big( \frac{{\bf l v}}{{\bf l v }- i \Gamma_l}\frac{\partial
n_0}{\partial \epsilon} \Big) \ ,
\eea
with
\beq
G(\omega,k)=\int \frac{d^3p}{(2 \pi)^3}\frac{\delta f(0,k,p)}{{\bf k}{\bf
v} - \omega} \ 
\eeq
and 
\beq
\label{ep}
\epsilon(\omega,k)=1-\frac{\partial U_k}{\partial \rho} \int \frac{d^3 p}{(2
\pi)^3}\frac{{\bf kv}}{{\bf  kv}-\omega}\frac{\partial n_0}{\partial
\epsilon} \ .
\eeq

The time dependence of the density perturbation $\delta \rho$
 can be found using the inverse transform~:
\beq
\label{inv}
\delta \rho(t,k)=  \int_{-\infty + i \sigma}^{\infty + i \sigma}
\frac{d \omega}{2 \pi } \delta \rho(\omega,k) \ e^{-i\omega t} \ ,
\eeq
the integration path in the inverse Fourier transform laying above
any singularity of the integrand. 

As a first case we take only one unstable mode~:
\beq
\rho^{+}(k)=(2\pi)^3 \delta^3(k-k_0) \delta A_{k_0} \rho \ .
\eeq
The nonlinear term is nonzero only for the mode with the doubled
wave-vector $2 k_0$~:
\bea
\delta \rho(\omega, k)  & = & \frac{-(2 \pi)^3 \delta^3(k-2k_0)
\delta A_{k_0}^2 \rho^2}{\epsilon(\omega,k)} \frac{1}{i \omega
+2\Gamma_{k_0} }
\nonumber \\  & & \Bigg[
  \frac{1}{8 \pi^2} 
\frac{\delta^2 U_{2k_0}}{\delta \rho^2} \int p^2 dp 
\left( 2 - \frac{\omega}{2 k_0 v}\ln\left(\frac{\omega+2k_0 v}{\omega
-2 k_0 v} \right) \right)  
 \frac{\partial n_0}{\partial \epsilon} 
\nonumber \\  & & + \frac{k_0^2}{4 \pi^2}  \left(
 \frac{\delta
U_{k_0}}{\delta \rho} \right)^2   \int p^2 dp \Bigg( \frac{-2 i
\omega}{ (4(k_0v)^2-\omega^2)(i\omega+2\Gamma_{k_0})}
\nonumber \\
& & -
\frac{2\Gamma_{k_0}}{k_0 v (\omega -
2i\Gamma_{k_0})^2}\arctan\left(\frac{k_0v}{\Gamma_{k_0}}\right)
\nonumber \\ & &
+ \frac{i \Gamma_{k_0}}{k_0v(\omega-2i\Gamma_{k_0})^2}
\ln\left(\frac{\omega+2k_0 v}{\omega
-2 k_0 v}\right)  \Bigg) \frac{\partial n_0}{\partial \epsilon}  \Bigg]
\eea
The integral (\ref{inv}) can be calculated closing the
integration path in the lower half-plane.
Its value is determined by the singularities of the integrand. The
inverse susceptibility $1/\epsilon(\omega,k)$ has two poles
corresponding the solutions of the dispersion relation~:
\beq
\label{disp}
\epsilon(\omega,k)|_k=0 \ .
\eeq
The integrand also has a pole at $\omega=2i\Gamma_{k_0}$ and a cut on
the real axis in $\omega$ \cite{ja}. The contribution from the pole at 
$2i\Gamma_{k_0}$ is dominant at large times. Even if the the solution
of (\ref{disp}) is imaginary we always have $2\Gamma_{k_0} >
\Gamma_{2k_0}$. The cut contribution is always bounded and can be
neglected at large times in comparison to the growing components.
The result is~:
\bea
\label{gg}
\delta \rho(t, k)  & = & \frac{(2 \pi)^3 \delta^3(k-2k_0)
\delta A_{k_0}^2 \rho^2}{\epsilon(2i\Gamma_{k_0},2k_0)} 
e^{2\Gamma_{k_0}t}
\  \Bigg[   
\frac{\delta^2 U_{2k_0}}{\delta \rho^2} /\frac{2\delta U_{k_0}}
{\delta \rho} 
\nonumber \\  & & + \frac{k_0^2}{8 \pi^2}  \left(
 \frac{\delta
U_{k_0}}{\delta \rho} \right)^2   \int p^2 dp  \frac{-\Gamma_{k_0}}
{ ((k_0v)^2+\Gamma_{k_0}^2)^2}
 \frac{\partial n_0}{\partial \epsilon}  \Bigg] \nonumber \\
& = & (2\pi)^3 \delta^3(k-2k_0) F_{2k_0}(t) \ .
\eea

In Fig.1 is  shown the ratio of the amplitude $F_{2k_0}(T)$ 
of the mode
at $2k_0$ to $\rho$ at the instability time $T$ which is defined as
 $|\delta A_{k_0} \exp(\Gamma_{k_0}T)| \simeq 1$ \cite{c1,he}.
The result shows that the long wavelength modes are always
nonlinear. Before the time the instability drives the mode $k_0$ to
large values the mode with double wave-vector becomes larger. However,
the amplitude of 
the linear modes with large wave-vectors ($k>.4 fm^{-1}$)
 is large ($|\delta A_k \exp(\Gamma_k T )|\simeq \rho$)
before   the nonlinearly
driven mode at $2k_0$ becomes important. 
In particular in our case the most unstable mode
is in the region of wave-vectors, where the linear regime is valid up
to the instability time $T$. Similar observation were made in the
2-Dimensional numerical solutions of the Vlasov equation,
 the small wave-vector modes become
nonlinear at smaller amplitude than the large wave-vector modes \cite{ch2}.
In Fig.2 is shown the growth time for the amplitude of the mode to
reach $\rho$. We see that for the long-wavelength the growth time for
the nonlinear mode at $2k_0$ is smaller than the growth time for the 
linear mode at $k_0$. The unphysically long values of the growth times
depends of course on the initial perturbation chosen
 (here $\delta A=1/50$).

The nonlinearities become extremely important for long wave-length (Fig.1).
However in this limit one should take into account both the pole at 
$2i\Gamma_{k_0}$ and at $i\Gamma_{2k_0}$, since the two poles merge in 
that limit ($\Gamma_k \sim k$). The time dependence is now different~:
\bea
\label{pop}
\delta \rho(t, k)  & = & \frac{(2 \pi)^3 \delta^3(k-2k_0)
\delta A_{k_0}^2
\rho^2}{i\epsilon^{'}(\omega,2k_0)|_{\omega=2i\Gamma_{k_0}}} 
\ t \ e^{2\Gamma_{k_0}t} \ \Bigg[
\frac{\delta^2 U_{2k_0}}{\delta \rho^2} / \frac{2 \delta U_{k_0}}{\delta 
\rho}
\nonumber \\  & & + \frac{k_0^2}{8 \pi^2}  \left(
 \frac{\delta
U_{k_0}}{\delta \rho} \right)^2   \int p^2 dp  \frac{-\Gamma_{k_0}}
{ ((k_0v)^2+\Gamma_{k_0}^2)^2}
 \frac{\partial n_0}{\partial \epsilon}  \Bigg] 
\nonumber \\ & = & (2\pi)^3 \delta^3(k-2k_0)
 L_{k_0} \rho \  \delta A_{k_0}^2  \ t \ e^{2\Gamma t} \ .
\eea
Now, the amplitude $F_{2k_0}(T)$
 at the instability time $T$ depends on the value of the initial
perturbation $\delta A_{k_0}$~:
\beq 
F_{2k_0}(T)=-\frac{L_{k_0}}{\Gamma_{k_0}}\ln(|\delta A_{k_0}|) \ \rho \ .
\eeq
This shows that it is not the same to use larger initial fluctuations
evolved on shorter time to using small initial fluctuations evolved
for longer times. Although, the linear evolution would give the same
result $|\delta A \exp(\Gamma T)|=1$, the nonlinear modes at double
wave-vector would have a different strength. 
In long wavelength 
limit the 
procedure  \cite{os} of putting instead of the  noise term in the 
Vlasov equation a
stronger  initial  perturbation gives different results due to
nonlinearities of the Vlasov equation. 
However, the most unstable modes 
are behaving according to eq. (\ref{gg}). Moreover, the numerical
coefficient is such that the nonlinearities have no time to built up
until the amplitude of the  most unstable modes becomes  of the
 order $\rho$.
The results from the 
correct limiting expression for the
limit $\Gamma(k)\sim k$ (eq. \ref{pop}) is also shown in Fig.2. The correct
dependence of the growth time $T_{growth} \sim 1/k$  in the long wavelength
limit is recovered.

Another  case is  when two unstable modes develop. We shall
study the case  when the wave-vectors of the unstable modes are in 
opposite direction~:
\beq
\rho^{+}(k)=(2\pi)^3  \left(\delta^3(k-l) \delta A_l
+  \delta^3(k+k_0-l) \delta A_{k_0-l} \right) \ \rho \ ,
\eeq
with $k_0<l$ and $l$ corresponding to a strongly growing mode.
Besides the contributions at doubled wave-vector $2l$ and $2(k_0-l)$
we find a  nonlinear mode  located at $k_0$. The mechanism responsible 
for the growth of the mode at $k_0$ is
different from the simple wave-vector doubling. As previously the
dominant mode is given by the pole of (\ref{lll}) at
$i(\Gamma_l+\Gamma_{k_0-l})$, leading to the behavior~:
\beq
\label{iii}
F_{k_0}(t) \sim \delta A_{k_0-l} \delta A_l
e^{(\Gamma_l+\Gamma_{k_0-l})t} \ .
\eeq
We do not quote explicitly the lengthy formula for the coefficient,
given by  residue at the pole of (\ref{lll}).
In Fig.3 is shown the ratio $|F_{k_0}(T)/F_{k_0-l}(T)|$ at the time $T$
when the amplitude of the most unstable mode $l\sim.7fm^{-1}$ reaches $\rho$.
For $|k_0-l|<.3fm^{-1}$ the linear mode at $k_0-l$ is weaker than the
nonlinear mode at $k_0 > .3fm^{-1}$. Similar nonlinear effect were observed
in numerical simulations \cite{ch1}. In the case of a strongly excited
mode at relatively small wave-vector and a weakly excited  mode with strong
growth rate several modes at intermediate wave-vectors appear, leading
to strong nonlinearities. Of course if the most unstable mode $l$ is 
strongly excited  it will dominate both the linear mode at $k_0-l$ and the
nonlinear mode at $k_0$. When $k_0 \rightarrow l$ the contribution
from the pole of (\ref{lll}) at
$\Gamma_{k_0}$ cannot be neglected. This changes the behavior of the 
time dependence of the nonlinear mode from (\ref{iii}) to  a form
analogous to (\ref{pop}). Thus the singularity in Fig.3 when
$k_0\rightarrow l=.7fm^{-1}$ is spurious.

In summary, we have studied the importance of the nonlinearities in the 
Vlasov equation for the development of the spinodal instabilities.
 The mechanism of wave-vector doubling is important only in the regime 
of wave-vector $k<.35 fm^{-1}$ (this value is determined mainly by the 
range of the interaction). For the most unstable modes the
nonlinear effects at double wave-vector are small.
For the case of two exponentially growing modes with wave-vectors in
opposite direction, we have found that
nonlinear modes at intermediate wave-vectors can be excited.
Thus, a small excitation
 of a rapidly growing mode $\sim .7fm^{-1}$ together with a
strong excitation of the mode $-.3fm^{-1}$ leads to strong nonlinear
effects at wave-vectors $\sim.4fm^{-1}$.
If for a given wave-vector the nonlinearities are found to be
important, then the perturbative approach leading from (\ref{nn}) to
(\ref{p2})
becomes invalid at some time smaller than the instability time.
From the one and two mode cases studied in this letter any initial
one-dimensional perturbation of the homogeneous solution can be
constructed. In  the second order in $\delta f(t,k,p)$ the
corresponding nonlinearities from mode-mode coupling can be calculated.
Those estimates explain the observations made in numerical simulations 
of the 2-Dimensional
Vlasov equation \cite{ch1,ch2} and are the first 
estimate of nonlinear effects in 3-Dimensions.

\vspace{5mm}

\noindent
{\bf Acknowledgments} \\
The author wishes to thank  for the hospitality extended to him
by the YITP.
 
%\newpage

\newpage

% {\bf Figure captions} \\
 
\begin{figure}
\begin{center}
\epsfig{file=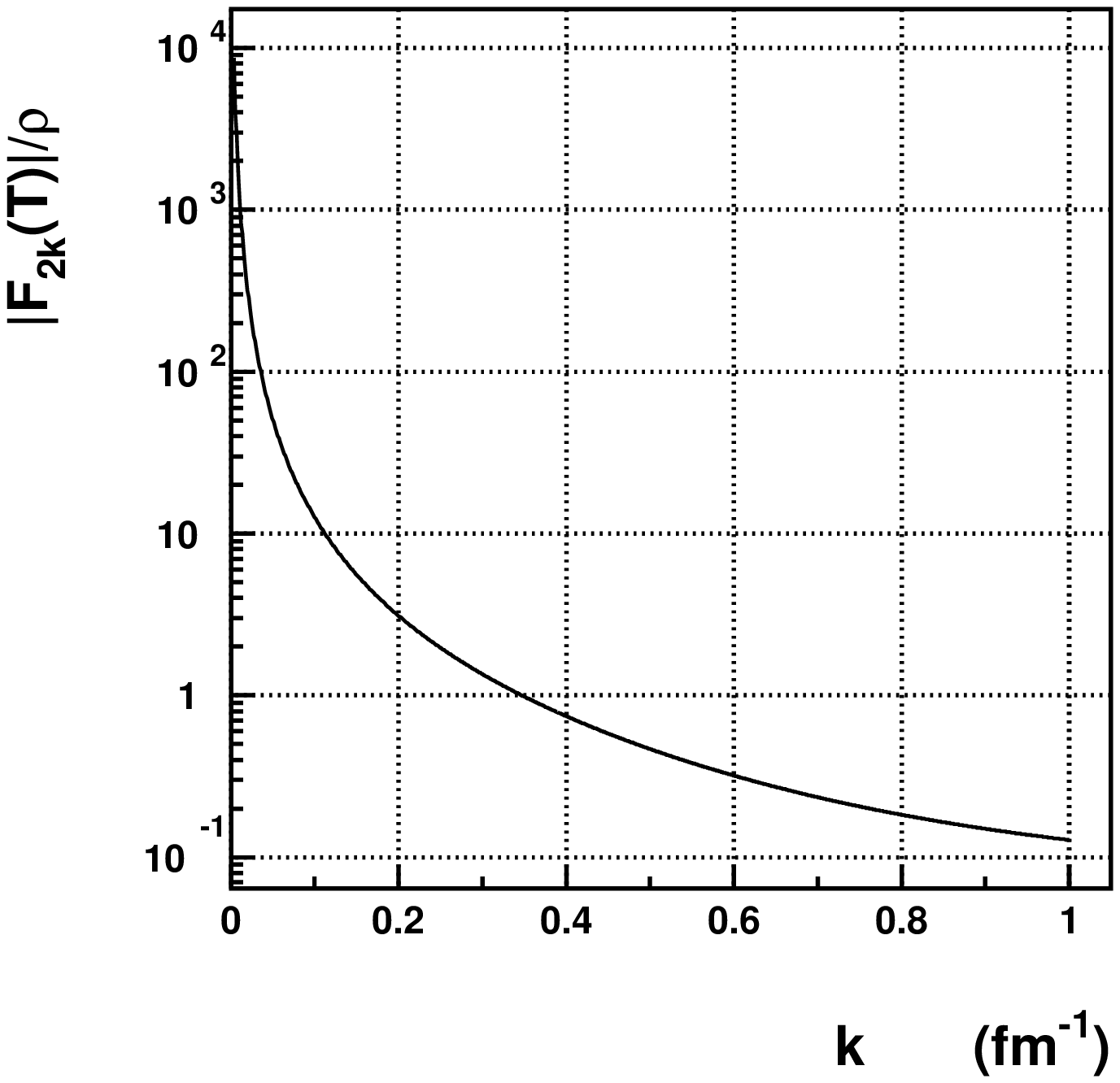,width=0.8\textwidth}
\vspace{0.5cm}
\end{center}

{\bf Fig.1} \\
The ratio of the amplitude $|F_{2k}(T)|$ of the nonlinear mode at
double wave-vector  $2k$ to the
average density $\rho$ as a function of the wave-vector at the
instability time $T=\ln(1/|\delta A_k|)/\Gamma_k$.
The calculation is done at zero temperature with the 
 parameters of the interaction  $A=-356MeV$, $B=303MeV$,
$\sigma=7/6$, $\rho=\rho_0/3$ and the range of the Gaussian equal to $.9fm$.
\end{figure}

\newpage
\begin{figure}
\begin{center}
\epsfig{file=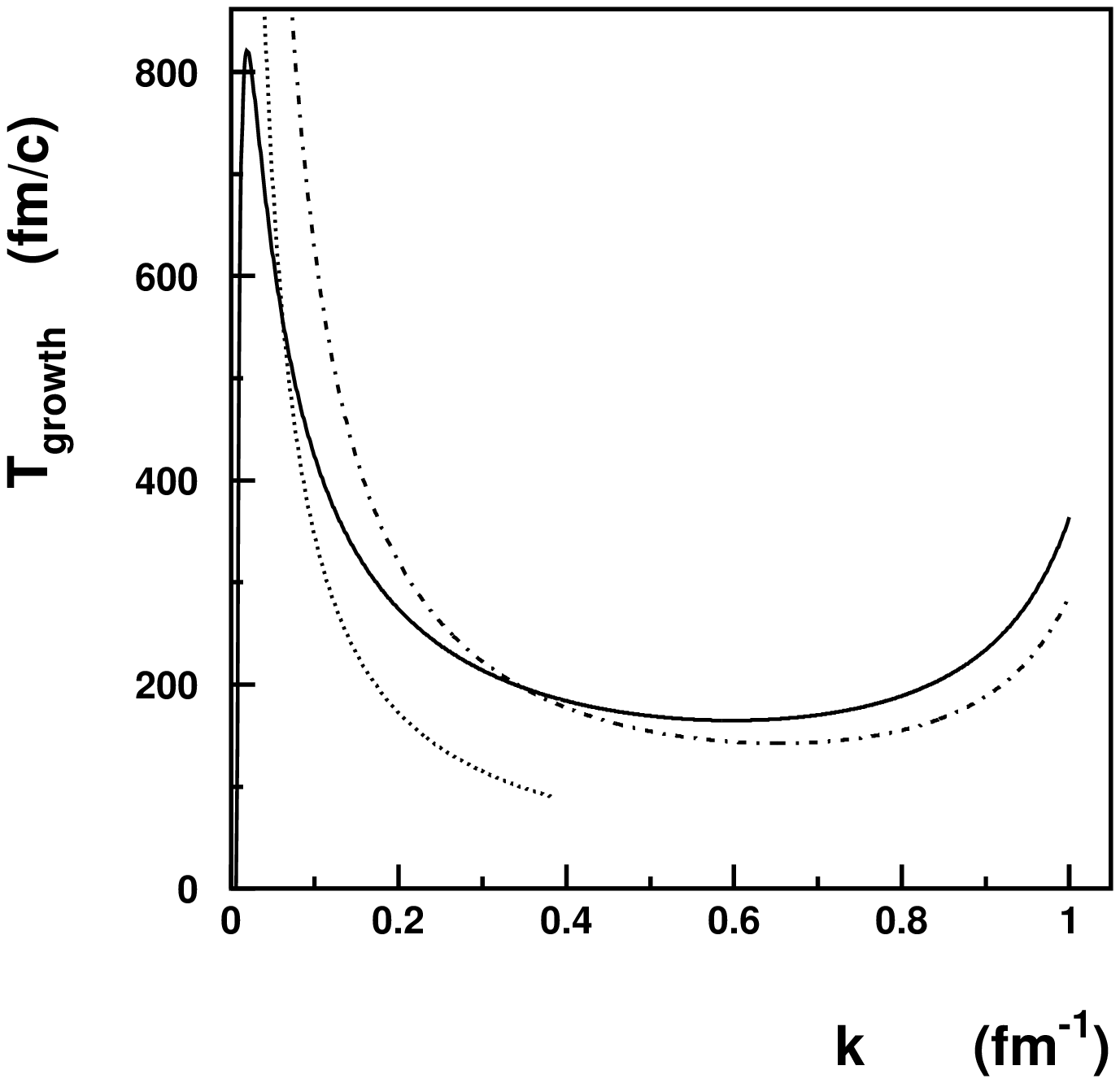,width=0.8\textwidth}
\vspace{0.5cm}
\end{center}

{\bf Fig.2} \\
The growth time $T_{growth}$ for the amplitude of a mode to reach the average
density $\rho$ with $\delta A_k=1/50$, as a function of the wave-vector. 
The solid line is for the nonlinear mode at $2k$, the 
dashed-dotted line is for the linear mode at $k$ and the dotted line
is for the nonlinear mode at $2k$ in the regime $\Gamma_k \sim k$.
The parameters used are the same as in Fig.1.
\end{figure}

\newpage
\begin{figure}
\begin{center}
\epsfig{file=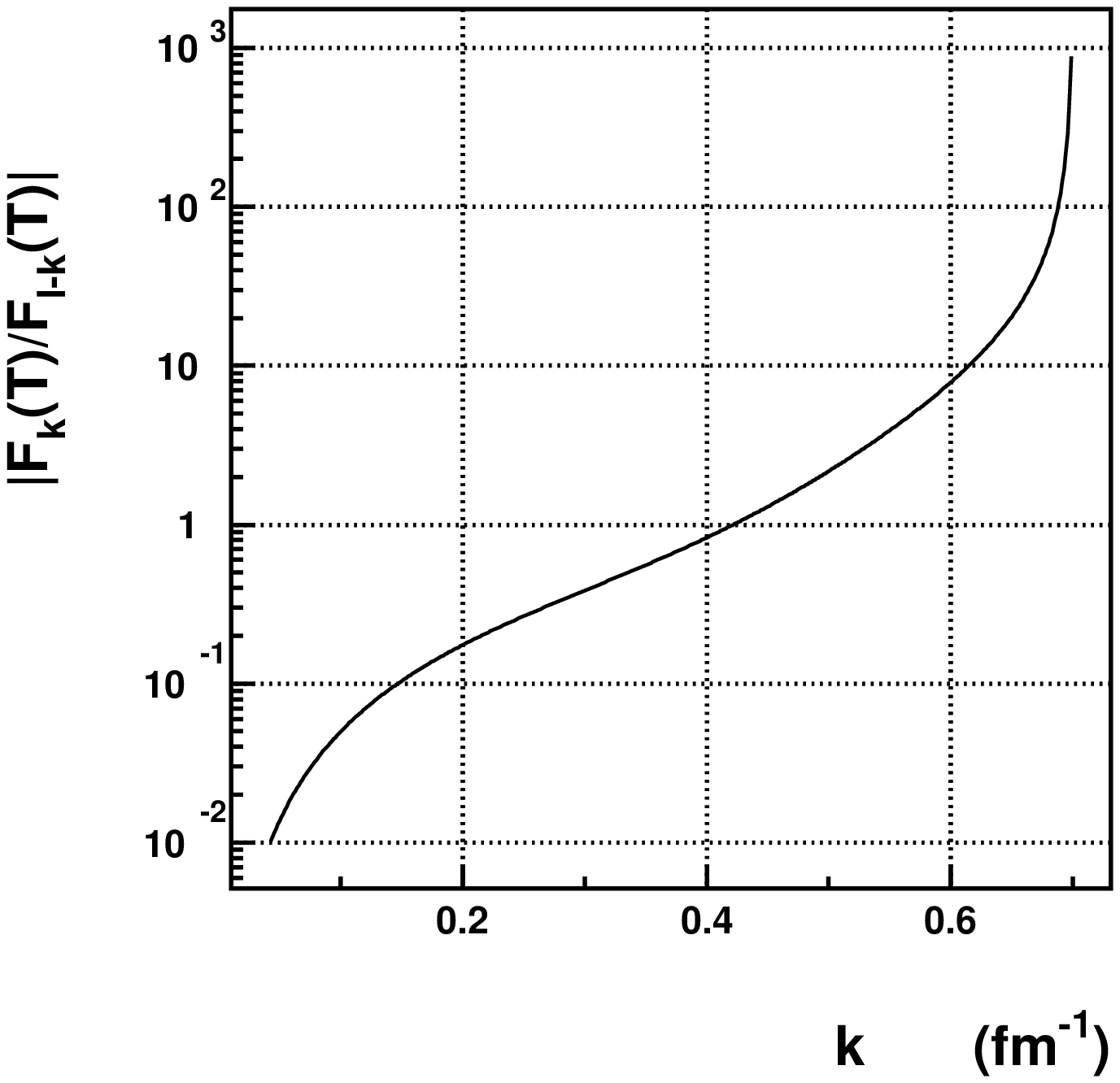,width=0.8\textwidth}
\vspace{0.5cm} 
\end{center}

{\bf Fig.3} \\
The ratio of the amplitude $F_{k}(T)$ of the nonlinear mode at $k$
 to the amplitude $F_{k-l}(T)$ of the linear mode at  $k-l$ as 
a function of the wave-vector at the
time $T=ln(1/|\delta A_l|)/\Gamma_l$  ($l=.7fm^{-1}$), i.e.
 when the  amplitude of the
other linear mode reaches the average density.
The parameters used are the same as in Fig.1.

\end{figure}

\end{document}